\documentclass[aps,prb,twocolumn,superscriptaddress]{revtex4-1}
\usepackage{color}
\usepackage{graphicx}
\usepackage{cancel}
\usepackage{amsfonts}

\usepackage[usenames,dvipsnames]{xcolor}
\definecolor{terracotta}{rgb}{0.89, 0.45, 0.36}

\usepackage{soul}

\begin{document}

\title{Charge and spin conductivity of a two-dimensional electron gas
with random Rashba interaction}

\author{S. Kud{\l }a}
\affiliation{Department of Physics and Medical Engineering, Rzesz\'ow University of Technology,
Al.~Powsta\'nc\'ow Warszawy 6, 35-959 Rzesz\'ow, Poland}

\author{A. Dyrda\l}
\affiliation{Institut f{\"u}r Physik, Martin-Luther Universit{\"a}t Halle-Wittenberg, D-06099 Halle, Germany}
\affiliation{Faculty of Physics, Adam Mickiewicz University, ul. Umultowska 85, 61-614 Pozna\'n, Poland}

\author{V. K. Dugaev}
\affiliation{Department of Physics and Medical Engineering, Rzesz\'ow University of Technology,
Al.~Powsta\'nc\'ow Warszawy 6, 35-959 Rzesz\'ow, Poland}

\author{E. Ya. Sherman}
\affiliation{Department of Physical Chemistry, Universidad del Pa\'is Vasco UPV-EHU, 48080, Bilbao, Spain}
\affiliation{IKERBASQUE Basque Foundation for Science, Alameda Urquijo 36-5, 48011, Bilbao, Bizkaia, Spain}

\author{J. Barna\'s}
\affiliation{Faculty of Physics, Adam Mickiewicz University, ul. Umultowska 85, 61-614 Pozna\'n, Poland}
\affiliation{Institute of Molecular Physics, Polish Academy of Sciences, ul. M. Smoluchowskiego 17, 60-179 Pozna\'n, Poland}

\begin{abstract}
We calculate the transport relaxation time $\tau _{\rm tr}$ and spin transport
relaxation time $\tau _{s,{\rm tr}}$ for
a two-dimensional  electron gas with spatially fluctuating Rashba
spin-orbit  interaction.
These relaxation times determine the electrical  and  spin
conductivity of the two-dimensional system, respectively. It is shown that the transport relaxation
time $\tau _{\rm tr}$ is a nonmonotonic function of
electron energy $\varepsilon $, whereas the spin transport relaxation time $\tau _{s,{\rm tr}}$  decreases
with increasing $\varepsilon $, similarly  to the conventional electron relaxation time $\tau$
that characterizes the decay of an electron state corresponding to  certain values of the momentum and spin.
Such a behavior of the relaxation
times leads to unusual temperature dependence of the electrical and spin conductivity.
\end{abstract}

\date{\today }

\maketitle

\section{Introduction}

It is well known that Rashba spin-orbit interaction in two-dimensional (2D)
electron systems appears due to asymmetry
in confining potentials on both sides of the corresponding
heterostructure.\cite{rashba60,bychkov84,nitta97,pfeffer99,koga02}
Such an interaction leads to various spin-orbit
related effects, like spin Hall effect, current-induced spin polarization, spin-orbit torque, and others.
\cite{zutic04,sinova04,engel07,fabian07,dyakonov08,nagaosa10,sinova15}
These phenomena have been extensively studied in recent years.
However, the Rashba spin-orbit coupling should disappear in two-dimensional electron systems that exhibit symmetry with
respect to reflection in the 2D plane. In other words, the corresponding  coupling
constant~\cite{bychkov84} vanishes then by symmetry reasons. An example of such a system
is a symmetric semiconductor quantum well.

However, even though the symmetry precludes the presence of a uniform Rashba interaction, such a symmetry does not exclude
the existence of spatially fluctuating Rashba field with the corresponding mean value equal to zero. It was
already shown in detail how the spatially fluctuating Rashba field can appear due to a deviation from homogeneity
of the doping impurity distribution in the vicinity of a semiconductor quantum well.\cite{sherman03,glazov05,glazov10}
The main characteristics of the random Rashba coupling in such systems have been studied theoretically
in Refs.~[\onlinecite{sherman03,glazov05,glazov10}]. Moreover, recent experiments on scanning tunneling spectroscopy
of InSb surfaces allowed to obtain a pattern of the Rashba
coupling with the $\sim 1$ nm spatial resolution, which revealed strong randomness of this coupling.\cite{bindel16}

It was also demonstrated that the fluctuating Rashba field can induce a variety of experimentally
observable effects. For example, the spin Hall conductivity in 2D system with homogeneous Rashba interaction vanishes in the presence of spin-independent disorder~\cite{sinova04,engel07,sinova15}, but it is robust to scattering on impurities in the presence of spatially fluctuating Rashba field~\cite{dugaev10,dyrdal2012}. This nonzero value of
spin-Hall conductivity not only  agrees with the analysis based on the SU(2) symmetry of the
spin-orbit coupling \cite{raimondi2012} and
detailed numerical calculations \cite{moca2008},  but also can be considered as a mechanism of the spin-charge conversion in 2D systems.
\cite{huang2017}  Furthermore,  the fluctuating spin-orbit
interaction is  responsible for spin relaxation.\cite{glazov05,dugaev09,glazov10} For instance, it is possible
that the electron spin relaxation in a free standing graphene is related to the fluctuating Rashba field
arising from rippled graphene sheet,\cite{glazov10,dugaev11} random impurity-induced
spin-orbit coupling,\cite{Zhang2012} or strong effects of the randomness
introduced by the corrugation.\cite{vicent2017} In addition, the random Rashba
fields play an important role in transport properties of the edge states in topological
insulators \cite{strom2010,kimme2016,dolcini2017}, and also can be crucially important in systems with very
strong spin-orbit coupling.\cite{brosco2017}

In this work we consider the effect of spatially fluctuating Rashba field on the charge and spin conductivity
of a 2D electron gas. We assume that the mechanism of electron scattering from the Rashba field is dominant
for both momentum
and spin relaxation of electrons, which may happen  at very low density of impurities and defects.
On the other hand, the contribution of any other scattering mechanism can be taken into account effectively
by assuming a certain relaxation time (e.g. due to impurities and defects), and then by using the
Matthiessen rule to add the rates related to different relaxation mechanisms.

In section 2 we describe the model Hamiltonian assumed to describe the system with spatially fluctuating Rashba field, and also introduce Hamiltonian describing interaction of the system with external electromagnetic field.  Relaxation time is calculated in section
3, whereas the vertex function is derived in section 4. Electrical conductivity is calculated and discussed in section 5. In turn, in section 6 we calculate the spin current and spin conductivity. Final conclusions are in section 7.

\section{Model}

To describe the 2D electron system with fluctuating Rashba field we use the following Hamiltonian:
\begin{equation}
\label{1}
\hat{H}=\hat{H}_{0}+\hat{H}^{({\rm so})},
\end{equation}
where the first term corresponds to the kinetic energy of 2D electrons with parabolic energy spectrum,
\begin{equation}
\label{2}
\hat{H}_{0}=-\frac{\hbar ^2\nabla ^2}{2m^{*}},
\end{equation}
while the term $\hat{H}^{({\rm so})}$ stands for the random Rashba spin-orbit coupling,
\begin{equation}
\label{3}
\hat{H}^{({\rm so})}
=-\frac{i}{2}\, \sigma _{x}\left\{ \nabla _{y},\,\lambda (\mathbf{r})\right\}
+\frac{i}{2}\, \sigma _{y}\left\{ \nabla _{x},\,\lambda (\mathbf{r})\right\} .
\end{equation}
Here, $\sigma_x$ and $\sigma_y$ are the Pauli matrices acting in the spin space, whereas $\lambda ({\bf r})$ is the Rashba parameter that varies randomly in the 2D space. We assume that the average value of this parameter vanishes,
$\langle \lambda ({\bf r})\rangle =0$, so that the random field is characterized by the
correlator $\langle \lambda ({\bf r})\, \lambda ({\bf r'})\rangle $.\cite{dugaev09,glazov10}
The matrix elements of the Rashba spin-orbit interaction (3) in the basis of the eigenfunctions of
Hamiltonian $\hat{H}_{0}$  are
\begin{eqnarray}
\label{4}
H^{({\rm so})}_{\bf kk'}=\frac{\lambda_{\bf kk'}}{2}\left[\sigma_x(k_y+k_y')-\sigma_y(k_x+k_x')\right] .
\end{eqnarray}

Now we assume that the system is in an external electromagnetic field described by a vector potential
${\bf A}(t)={\bf A}_0\, e^{-i\omega t}$.
To find  Hamiltonian which describes interaction of the system under consideration with the electromagnetic field, we make the
replacement: $\mathbf{k} \to \mathbf{k}-e\mathbf{A}/\hbar c$. Accordingly,
the corresponding spin-orbit dependent part of the interaction
with electromagnetic field can be written as
\begin{eqnarray}
\label{5}
H^{({\rm so})-A}_{\bf kk'}=-\frac{e\lambda_{\bf kk'}}{c}\left(\sigma_xA_y-\sigma_yA_x\right) .
\end{eqnarray}
When taking into account also the interaction of free electrons with the electromagnetic field,
the total Hamiltonian describing coupling of the system  to the
electromagnetic field can be written in the following form:
\begin{eqnarray}
\label{6}
H^{(A)}_{\bf kk'}=
-\frac{e\hbar {\bf k}\cdot {\bf A}}{m^{*}c}\, \delta _{\bf kk'}
+\frac{e^2A^2}{2m^{*}c^2}\, \delta _{\bf kk'}
\nonumber \\
-\frac{e\lambda_{\bf kk'}}{\hbar c}\left(\sigma_xA_y-\sigma_yA_x\right).
\end{eqnarray}
Here, the first and second terms are the usual kinetic and diamagnetic contributions. The third term, in turn,
takes into account the coupling mediated by the spin-orbit interaction and corresponds to the anomalous
spin-dependent velocity in the form of the commutator $i[\hat{H}^{({\rm so})},\mathbf{r}]/\hbar$.

Without loss of generality, we assume that the vector potential ${\bf A}$ is along the $x-$axis,
and calculate the current
flowing along this axis. The corresponding matrix elements of the charge current operator can be then written in the form
\begin{eqnarray}
\label{7}
&& j_{x,{\bf kk'}}=-c\, \frac{\partial H^A_{\bf kk'}}{\partial A_x} \\
&&\quad\,=\frac{e\hbar }{m^{*}}\, \left( k_x-\frac{eA_x}{\hbar c}\right) \delta _{\bf kk'}
-e\lambda_{\bf kk'}\sigma_y . \nonumber
\end{eqnarray}
To calculate the electric current flowing in the system we will use the standard Kubo formalism and Green function technique in the loop approximation, with a renormalized vertex function.\cite{agd,mahan} To avoid issues related to  electron localization, all the calculations will be performed assuming that scattering from the random Rashba field is weak. This scattering gives rise to
a slow  relaxation of electron states described by certain
momentum and spin.
The other
effect due to fluctuating Rashba field is related to its correction to the current vertex,  similar to the
impurity-induced correction. \cite{agd}
This leads to the substitution of the bare current vertex $j_{x,{\bf kk'}}$ by its
renormalized counterpart $J_{x,{\bf kk'}}$.
Note that the vertex correction does not vanish in the limit of very weak scattering by the
fluctuating field since the relative correction to the bare vertex is of the order of unity.
\cite{agd}

Assuming the weak scattering regime, we restrict ourselves to the first term in the right-hand-side of
Eq.~(\ref{7}). Correspondingly, we do not take into account the last (anomalous velocity-related) term in Eq.~(\ref{6}) related to the
spin-orbit induced interaction with electromagnetic field. The above-mentioned terms lead to a negligibly small
correction to the calculated conductivity. Indeed, the main contribution to the conductivity is
of the order of $(e^2/\hbar )\, (\varepsilon _F\tau _{\rm tr}/\hbar )$, where $\varepsilon _F$ is the Fermi energy and
$\tau _{\rm tr}$ is the transport relaxation time,
whereas the correction related to the  anomalous velocity is $\sim e^2/\hbar $. Below we use the units with $\hbar\equiv 1$
and restore $\hbar$ in the numerical calculations.

Before calculating the electrical current and the corresponding conductivity, we need to find the relaxation time and the vertex function. These will be derived in the following two sections.

\section{Relaxation times}

Now we calculate the relaxation time due to scattering on the fluctuating spin-orbit  Rashba field.
Since the fluctuations are assumed to be small, they can be
considered in terms of the perturbation theory.
The Green function for electrons in 2D electron gas with disorder can be written in the following  general form:
\begin{eqnarray}
\label{8}
G^{R,A}_{\varepsilon ,{\bf k}}=\frac1{\varepsilon -\varepsilon _{k}\pm i/2\tau},
\end{eqnarray}
where $\varepsilon _k=k^2/2m^{*}$, the energy  $\varepsilon $ is measured from the bottom of the electron energy band,  and $1/2\tau$ is the relaxation rate  due to scattering
from the fluctuations of the spin-orbit field. We assume that there are no other scattering centers in the system (like impurities or
defects) which would lead to decay of the electron state with the momentum ${\bf k}$.

In the Born approximation, the self energy due to scattering from fluctuating spin-orbit field can be calculated from the formula,
\begin{eqnarray}
\label{9}
&&\Sigma ^{R}(\varepsilon ,k)=\\
&&\quad\,-\frac{i\pi }4 \sum _{\bf k'}|\lambda _{\bf kk'}|^2\,
\left(k^2+k'^2+2{\bf k}\cdot {\bf k'}\right)\,
\delta (\varepsilon -\varepsilon _{k'}). \nonumber
\end{eqnarray}
In the following we assume the disorder correlator in the form\cite{dugaev09,glazov10},
\begin{eqnarray}
\label{10}
\left< |\lambda _{\bf kk'}|^2\right> =2\pi \left< \lambda ^2\right> R^2e^{-|{\bf k-k'}|R} ,
\end{eqnarray}
where $R$ is the correlation radius of the spatial fluctuations, while the parameter
$\langle \lambda ^2\rangle $ characterizes amplitude of these fluctuations.
Then, upon averaging over static disorder we obtain from Eq.~(\ref{9})
the following expression for the relaxation rate,
$1/\tau =2{\rm Im}\, \Sigma ^R(\varepsilon _k, k)$:
\begin{equation}
\label{12}
\frac{1}{\tau}=\frac{1}{\tau_0}{R^2k^2}
\int _0^{ \pi }d\varphi \, e^{-2Rk\left|\sin\left(\varphi/2\right)\right|}
\left(1+\cos \varphi\right),
\end{equation}
where we introduced a constant $\tau_0$ for the time scale,
\begin{eqnarray}
\label{13}
\frac{1}{\tau_0}\equiv\left<\lambda^{2}\right> m^{*}.
\end{eqnarray}
The parameter $\tau_0$ has the physical meaning
of a characteristic spin rotation time of a particle with momentum $m^{*}\langle\lambda^2\rangle^{1/2}$
in a constant spin-orbit  field $\langle\lambda^2\rangle^{1/2}.$

\begin{figure}[h]
\centering
\includegraphics[width=0.99\columnwidth]{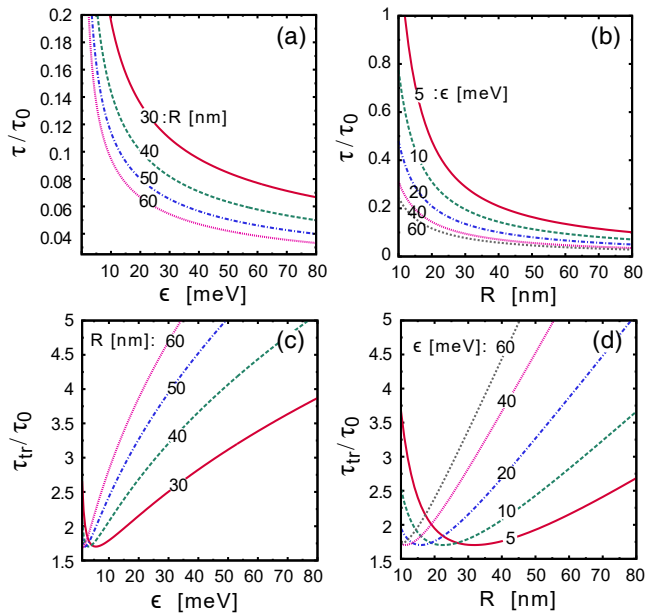}
\caption{The relaxation time $\tau/\tau_{0}$ (panels (a),(b)) and transport relaxation time $\tau_{\rm tr}/\tau_{0}$
(panels (c),(d)) (both normalized to $\tau_0$), shown as a function of energy for different values of the disorder correlation
length $R$ (panels (a) and (c)) and as a function of $R$ for different values of $\varepsilon$ (panels (b) and (d)).
Other parameters: electron effective mass $m = 0.03 m_{0}$}
\label{fig1}
\end{figure}

Dependence of the relaxation time $\tau $ on the electron energy, as determined by Eq.~(\ref{12}),
is presented in Fig.~\ref{fig1}(a). This figure shows that $\tau $ is divergent for
$\varepsilon\to 0$ and  decreases
with increasing $\varepsilon$.  The divergence is a consequence of the $k-$dependence in Eq.~(\ref{12}),
where electrons with large wavelengths, $2\pi/k\gg R$, do not {\it see} the short-range spin-orbit fluctuations.
The prefactor in Eq.~(\ref{12}) includes $k^2$ and thus goes to zero for $\varepsilon\to 0$. Note, such a divergence is removed when including scattering from impurities.
In turn, behavior of the relaxation time for $\varepsilon >0$
is a consequence of the interplay of increase in the prefactor and decrease in the exponential term under the integral.
As a result, the relaxation time $\tau$ as a function of $kR$ behaves as $\sim 1/\left(kR\right)^2$
at $kR\ll1$ and as $\sim 1/kR$ at $kR\gg1$. This behavior corresponds to Fig. \ref{fig1}(a) and to
Fig. \ref{fig1}(b), where the $R$-dependence of $\tau$ is presented. In addition, it should be noted
that  the time $\tau$ in our model becomes simultaneously also the spin relaxation time
because neither spin nor momentum are conserved in scattering from the Rashba field.

\section{Current vertex}

To calculate the electrical conductivity in the Kubo formalism we need the current vertex renormalized
by the fluctuating spin-orbit Rashba field.
As already mentioned above, we take the bare (unrenormalized) current vertex in the form
\begin{eqnarray}
\label{14}
j_x({\bf k})=\frac{e k_x}{m^{*}} .
\end{eqnarray}
Then, the ladder equation for the renormalized current vertex $J_x(\varepsilon ,\varepsilon ', {\bf k})$
takes the form
\begin{eqnarray}
\label{15}
&&J_x(\varepsilon ,\varepsilon ', {\bf k})
=j_x({\bf k})+\frac{1}{4} \sum _{\bf k'} \, J_x(\varepsilon ,\varepsilon ',{\bf k'})
|\lambda _{\bf kk'}|^2\,
\nonumber \\
&&\times[\sigma_x(k_y+k'_y)-\sigma _y(k_x+k'_x)]\,
G_{\varepsilon ,{\bf k'}}^A
\nonumber \\
&&\hspace{0.8cm}\times[\sigma _x(k_y+k'_y)-\sigma _y(k_x+k'_x)]\, G_{\varepsilon ',{\bf k'}}^R .
\end{eqnarray}
For brevity of notation, we will omit below $\varepsilon$ and $\varepsilon '$ in $J_x(\varepsilon ,\varepsilon ', {\bf k})$.
Then, using Eq.~(\ref{8}) for the Green's function, we obtain from Eq.(\ref{15})
the following equation for $J_x({\bf k})$:
\begin{eqnarray}
\label{16}
J_x({\bf k})&&=\frac{e k_x}{m^{*}}+\frac{i\pi }{4(\omega +i/\tau)} \sum _{\bf k'} \, J_x({\bf k'})
|\lambda _{\bf kk'}|^2\,
\nonumber \\
&&\hspace{-0.8cm}\times(k^2+k'^2+2{\bf k}\cdot {\bf k'})\,
[\delta (\varepsilon -\varepsilon _{k'})+\delta (\varepsilon '-\varepsilon _{k'})] ,\hskip0.3cm
\end{eqnarray}
where $\omega =\varepsilon '-\varepsilon $.

The detailed calculations, which include disorder averaging and the limit of $\omega\to\,0$
(see the Appendix A) lead
to the vertex function in the form
\begin{eqnarray}
\label{17}
J_x(\varepsilon ,\varepsilon ;{\bf k})=\frac{e k_x}{m^{*}}\, \frac{\tau _{\rm tr}}{\tau }\, ,
\end{eqnarray}
where $\varepsilon =k^2/2m^{*}$ and the transport relaxation time $\tau _{\rm tr}$ is given by the formula
\begin{eqnarray}
\label{18}
\frac{1}{\tau _{\rm tr}}=\frac{1}{\tau_{0}}R^2k^{2}
\int _0^{ \pi }d\varphi \, e^{-2Rk\left|\sin\left(\varphi/2\right)\right|}\, \sin ^2\varphi ,
\end{eqnarray}
with $\tau_{0}$ defined by Eq.~(\ref{13}).

Variation of the transport relaxation time $\tau_{\rm tr}$ with the energy and correlation
radius $R$ is shown in Figs.~\ref{fig1}(c) and \ref{fig1}(d). In the limit of small energy and small $R$,
behavior of $\tau_{\rm tr}$ is similar to that for $\tau$ since both these quantities show the
$1/\left(kR\right)^2$ divergence. In the opposite limit $kR\gg1$, the contribution from
scattering angles $\sim 1/kR$ dominates in the  relaxation rate $1/\tau$.
However, this small-angle scattering only weakly contributes to the $1/\tau_{\rm tr}$ rate.
As a result,  $\tau_{\rm tr}$ behaves as $\sim kR$ in the $kR\gg1$ limit, which leads to the increase in $\tau_{\rm tr}$
upon reaching the minima, as clearly seen  in Figs.~\ref{fig1}(c) and \ref{fig1}(d) (note, such minima are absent in Figs.~\ref{fig1}(a) and \ref{fig1}(b)).

\section{Electrical conductivity}

Having found the relevant relaxation rates and the vertex function, one can calculate the electrical conductivity.
Using the Matsubara technique for finite-temperature Green functions, with discrete frequencies $i\omega_m=2im\pi T$,  where $m$ is an integer number and $T$ is the temperature, we get the following expression for the charge current\cite{agd,mahan}:

\begin{eqnarray}
\label{18b}
j_{x}(i\omega _m)=-\frac{e A_{x}T}{m^{\ast}c}\sum _{n{\bf k}}J_x({\bf k};i\varepsilon _n, i\varepsilon _n+i\omega _m)
\nonumber \\ \times
G_{\bf k}(i\varepsilon _n+i\omega _m)\, k_x\sigma _0\, G_{\bf k}(i\varepsilon _n) .
\end{eqnarray}

\begin{figure}
	\includegraphics[width=0.99\columnwidth]{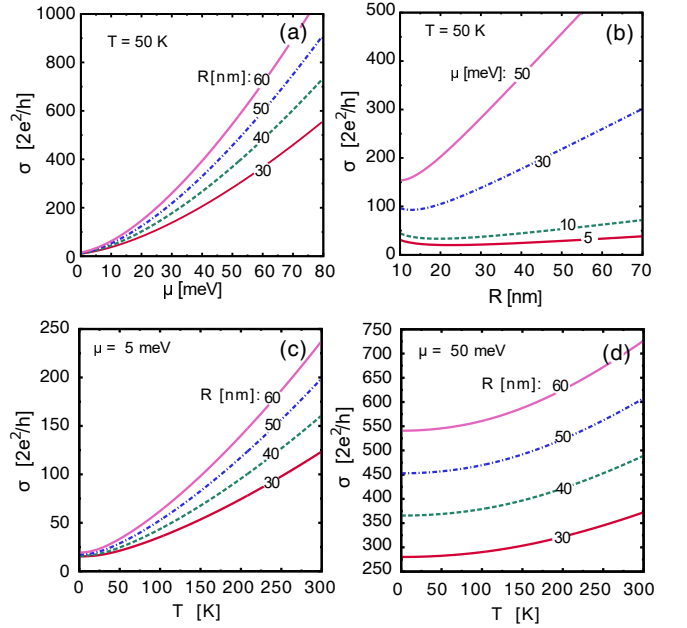}
	\caption{Electric conductivity as a function of chemical potential $\mu$ (a), correlation length $R$ (b),
	and temperature $T$ ((c), (d)).
	Other parameters: electron effective mass \cite{becker2010}  $m = 0.03 m_{0}$,  $\sqrt{\langle \lambda^{2}\rangle} = 0.15$ eV{\AA}, and
	the corresponding  $\tau_{0}=0.22$ ns, as follows from Eq. (\ref{13}).}
	\label{fig2}
\end{figure}

Upon analytical continuation to real frequencies, $i\omega _m\to \omega $, one obtains
\begin{eqnarray}
\label{20}
&j_x(\omega )=-\frac{eE_0}{2\pi \omega m^{*}}\, {\rm Tr} \int \frac{d^2{\bf k}}{(2\pi )^2}\, k_x \int _{-\infty }^\infty\,d\varepsilon\Big[ f(\varepsilon +\omega )-f(\varepsilon )\Big]\nonumber \\
&\times \,
J_x({\bf k},\varepsilon , \varepsilon +\omega )\, G^R_{{\bf k}}(\varepsilon +\omega)\, G^A_{{\bf k}}(\varepsilon),\hspace{0.5cm}
\end{eqnarray}
where the Fermi-Dirac distribution $f(\varepsilon)$ is given by
$1/\left[\exp\left((\varepsilon-\mu)/T\right)+1\right]$, with the chemical potential $\mu$.
Then, using Eq.~(\ref{17}) for the vertex function $J_x({\bf k},\varepsilon ,\varepsilon )$, one finds from Eq.(\ref{20}) the following formula for the static  ($\omega \to 0$) charge current:
 \begin{equation}
\label{21}
j_x=-\frac{e^{2}E_0}{4\pi ^2m^{*}}
\int _{0}^\infty  d\varepsilon f'(\varepsilon )\,   \,
{k^2}\,\tau _{\rm tr}.
\end{equation}
From this formula follows that the corresponding electrical conductivity $\sigma $
is determined by the transport relaxation time $\tau _{\rm tr}$ as
\begin{eqnarray}
\label{22}
\sigma = -\frac{e^2}{2\pi^2}
\int _{0}^\infty d\varepsilon \, \varepsilon f'(\varepsilon )\, \tau _{\rm tr}(\varepsilon )\, .
\end{eqnarray}

Numerical results for the electrical conductivity, obtained from Eq. (\ref{22}) with the transport time given by
Eq. (\ref{18}), are presented in Fig. \ref{fig2} for indicated parameters describing the system. The increase in
conductivity with increasing temperature and chemical potential results from increasing contribution of
electrons with higher momentum and from the modification of the total electron concentration.  All the dependencies presented in Fig.2
are related to the dependence of $\tau _{\rm tr}(\varepsilon)$ on the product $kR$,  see Eq. (\ref{18}).

\section{Spin current and spin conductivity}

To complete our considerations we analyze now the spin current flowing int the system.
The corresponding operator of spin current is defined as
\begin{eqnarray}
\label{23}
\hat{j}^z_x=\frac{k_x\sigma _z}{2m^{*}}\,
\end{eqnarray}
for transport of the $z$-component of spin polarization along the axis $x$.

\begin{figure}[h]
\centering
\includegraphics[width=0.99\columnwidth]{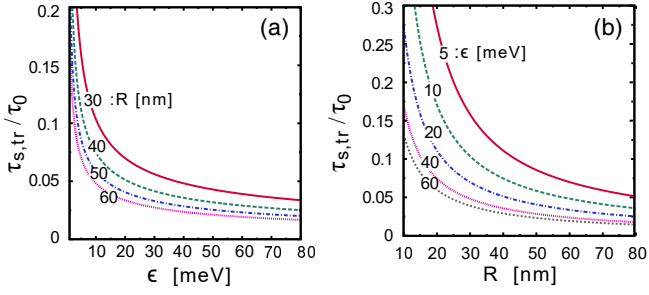}
\caption{{The normalized spin transport relaxation time,  $\tau_{\rm s,tr}/\tau_{0}$,
as function of energy $\epsilon$ for different values of the radius $R$ (a) and as a function of $R$ for
different values of energy (b). Other parameters: electron effective mass $m = 0.03 m_{0}$} }
\label{fig3}
\end{figure}

We assume that the spin current is generated by some spin vector potential $A^z_x(\omega )$,
so that the interaction of electrons with this field is described by the coupling
term \cite{scur}
\begin{eqnarray}
\label{24}
\hat{H}^{(A_s)}_{\bf k}=-\frac{k_x \sigma _zA^z_x}{2m^{*}c}\, .
\end{eqnarray}

\begin{figure}
	\includegraphics[width=0.99\columnwidth]{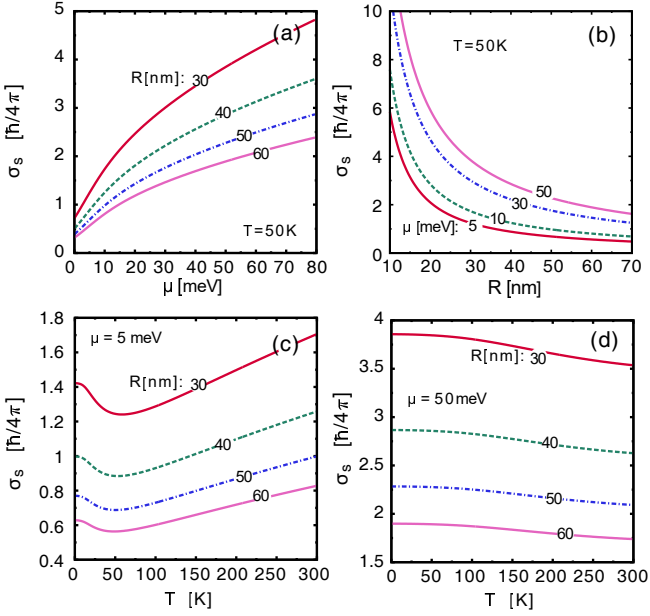}
	\caption{Spin conductivity as a function of the chemical potential $\mu$ (a), correlation length $R$ (b),
	and temperature $T$ ((c), (d)) for indicated parameters $R$, $T$ and $\mu$.
	Other parameters: electron effective mass $m = 0.03 m_{0}$, $\sqrt{\langle \lambda^{2}\rangle} = 0.15$ eV{\AA}.}
	\label{fig4}
\end{figure}

Using the Matsubara technique one arrives at the following formula for the spin current:
\begin{eqnarray}
\label{25}
j^z_x(i\omega _m)=-\frac{A^{z}_{x}T}{2m^\ast c}\sum _{n{\bf k}}J^z_x({\bf k};i\varepsilon _n, i\varepsilon _n+i\omega _m)
\nonumber \\ \times
G_{\bf k}(i\varepsilon _n+i\omega _m)\, k_x\sigma _z\, G_{\bf k}(i\varepsilon _n) ,
\end{eqnarray}
where $\varepsilon _n=(2n+1)\pi T$.
Then, upon analytical continuation, $i\omega _m\to \omega $, one can write
\begin{eqnarray}
\label{26}
&&\hspace{-0.5cm} j^z_x(\omega )=-\frac{E^z_{0x}}{4\pi \omega m^{*}}\, {\rm Tr} \int \frac{d^2{\bf k}}{(2\pi )^2}
\int _{-\infty }^\infty d\varepsilon \Big[ f(\varepsilon +\omega )-f(\varepsilon )\Big] \,
\nonumber \\
&&\times
J^z_x({\bf k},\varepsilon , \varepsilon +\omega )\, G^R_{{\bf k}}(\varepsilon +\omega )\,
k_x\sigma _z\, G^A_{{\bf k}}(\varepsilon),
\end{eqnarray}
where we introduced the spin electric field,
\begin{eqnarray}
\label{27}
{\bf E}^\alpha (t)\equiv {\bf E}^\alpha _0\, e^{-i\omega t}
=-\frac1{c} \frac{\partial {\bf A}^\alpha }{\partial t}.
\end{eqnarray}

To calculate the spin current from Eq.(25), we need to know the vertex function $J^z_x({\bf k})$. The corresponding equation for the spin current vertex reads
\begin{eqnarray}
\label{28}
&&{\bf J}^z({\bf k})=\frac{{\bf k}\sigma _z}{2m^{*}}
+\frac{i\pi }{4(\omega +i/\tau)} \sum_{\bf k'} {\bf J}^z({\bf k'})
|\lambda _{\bf kk'}|^2\,
\nonumber \\
&&\hspace{-0.5cm}\times(k^2+k'^2+2{\bf k}\cdot {\bf k'})\,
[\delta (\varepsilon -\varepsilon _{k'})+\delta (\varepsilon '-\varepsilon _{k'})] .
\end{eqnarray}
Solving this equation we find the following simple formula for the spin current:
\begin{eqnarray}
\label{29a}
J^{z}_{x}=\frac{k_x\sigma _z}{2m^{*}}\,\frac{\tau_{s,{\rm tr}}}{\tau},
\end{eqnarray}
where the spin transport relaxation time $\tau_{s,{\rm tr}}$ is given by the formula
\begin{equation}
\label{29b}
\frac{1}{\tau _{s,{\rm tr}}}=\frac{1}{\tau_{0}}R^2k^{2}
\int _0^{ \pi }d\varphi \, e^{-2Rk\left|\sin\left(\varphi/2\right)\right|}(1+\cos\varphi)^2.
\end{equation}

Dependence of the transport spin relaxation time on the electron energy and correlation radius
is presented in Fig. \ref{fig3}. Note that in contrast to $\tau_{\rm tr}$ presented in Figs. \ref{fig1}(c)
and \ref{fig1}(d),  ${\tau _{s,{\rm tr}}}$ does not increase at large $kR$, and behaves similarly
as the relaxation time $\tau$  presented in Figs. \ref{fig1}(a) and \ref{fig1}(b).
This difference is due to the fact that small-angle scattering
is essential for relaxation of spin current while it is  not essential for the relaxation of charge current.
As one can note when comparing Fig. \ref{fig3} and Fig. \ref{fig1},
the transport spin relaxation time is smaller than $\tau $ because it accounts for both effects of electron
scattering and spin relaxation.

Similarly to Eq. (\ref{22}), we obtain the spin conductivity
\begin{eqnarray}
\label{30}
\sigma _s= -\frac{1}{8\pi ^2}\int _{0}^\infty d\varepsilon\, \varepsilon f'(\varepsilon )\, \tau _{s,{\rm tr}}(\varepsilon )\, .
\end{eqnarray}
The numerical results for the spin conductivity, obtained from Eq. (\ref{30})
with the spin transport time given by Eq. (\ref{29b}), are presented in Fig. \ref{fig4}. These results can be accounted for in a similar way as the results for
electrical conductivity in Fig. \ref{fig2}.

\section{Conclusions}

In this paper we presented theoretical results on transport relaxation times $\tau _{\rm tr}$ and $\tau _{s,{\rm tr}}$,
responsible for the charge and spin conductivity
in a two-dimensional system, where the main mechanism of electron scattering is related to fluctuations
of the Rashba spin-orbit interaction. Although we considered an {\it extreme} situation, where the mean value of
the Rashba coupling is zero, this mechanism of electron scattering can be important
in a general case of $\langle\lambda({\bf r})\rangle \ne 0$, provided that
the spin-orbit coupling in the system is sufficiently strong.
It can be also related to defects in strongly spin-orbit coupled compounds such as transition-metal chalcogenides
or impurities at the surfaces or interfaces with a strong spin-orbit coupling.

The obtained transport relaxation times for the charge and spin current can be
essentially different from the times describing the electron momentum and spin relaxation.
The observed energy dependence of the transport time leads to nontrivial temperature
dependence of the conductivity which may increase with increasing temperature at a constant electron density.

\section*{Acknowledgements}
This work was supported by the National Science Center in Poland as a research project
No.~DEC-2012/06/M/ST3/00042. In addition, E.Y.S. acknowledges the support by the Spanish Ministry of Economy,
Industry, and Competitiveness and the European Regional
Development Fund FEDER through Grant No. FIS2015-67161-P (MINECO/FEDER),
and Grupos Consolidados UPV/EHU del Gobierno Vasco (IT-986-16). A.D. acknowledges the support by DFG through SFB762.

\appendix

\section{Calculation of the vertex function}

Here we present some details concerning calculation of the vertex function in the static
limit $\omega\rightarrow 0$ (which is  of our interest), see Eqs.~(\ref{15}) and (\ref{16}).
Since $J_x\sim k_x$,  Eq.~(\ref{16}) can be presented as an equation for the vector vertex ${\bf J}({\bf k})$
\begin{eqnarray}
\label{a1}
&&{\bf J}({\bf k})
=\frac{e{\bf k}}{m^{*}}+\frac{i\pi }{4(\omega +i/\tau)} \sum _{\bf k'}\, {\bf J}({\bf k'})
|\lambda _{\bf kk'}|^2\,
\nonumber \\
&&\times(k^2+k'^2+2{\bf k}\cdot {\bf k'})\,
[\delta (\varepsilon -\varepsilon _{k'})+\delta (\varepsilon '-\varepsilon _{k'})], \hskip0.5cm
\end{eqnarray}
and we can write ${\bf J}({\bf k})$ as
\begin{eqnarray}
\label{a2}
{\bf J}({\bf k})=\frac{e{\bf k}}{m^{*}}\; g(k) ,
\end{eqnarray}
where the scalar function $g(k)$ depends on the module of the vector ${\bf k}$ only. Then, from (\ref{a1}) and (\ref{a2})
we obtain in the limit $k\rightarrow\,k'$ the following equation for $g(k)$:
\begin{eqnarray}
\label{a3}
&&{\bf k}\, g(k)
={\bf k}+\frac{i\pi g(k)}{2(\omega +i/\tau)} \sum _{\bf k'}
|\lambda _{\bf kk'}|^2\, (k^2+k'^2+2{\bf k}\cdot {\bf k'})
\nonumber \\
&&\quad\times{\bf k'}\,\delta(\varepsilon -\varepsilon _{k'}).
\end{eqnarray}
The right-hand side of Eq.~(\ref{a3}) should be proportional to vector ${\bf k}$.
As a result, we find $g(k)$ from the following equation:
\begin{eqnarray}
&&g(k)=1+\frac{1}{\tau_{0}}R^2k^{2}{\tau}
\, g(k) \times \label{a6} \\
&&\int _0^{ \pi }d\varphi \,
e^{-2Rk\left|\sin(\varphi/2)\right|}
\cos^{2}\left(\varphi/2\right)\, \cos \varphi. \nonumber
\end{eqnarray}

Finally,  (\ref{a6}) yields
\begin{eqnarray}
\label{a8}
g(k)=\left(1-\frac{\tau}{\tau_{0}}{R^2k^{2}p(k)}\right)^{-1} ,
\end{eqnarray}
where we introduced the notation,
\begin{eqnarray}
\label{a9}
p(k)=\int _0^{ \pi }d\varphi \, e^{-2Rk\left|\sin\left(\varphi/2\right)\right|}\, (1+\cos \varphi )\, \cos \varphi .
\end{eqnarray}
Using the expression (\ref{12}) for $1/\tau$ we can write $g(k)$ in the following simple form
\begin{eqnarray}
\label{a10}
g(k)=\tau _{\rm tr}/\tau,
\end{eqnarray}
where the transport relaxation time $\tau _{\rm tr}$ is determined as:
\begin{eqnarray}
\label{a11}
\frac{1}{\tau _{\rm tr}}=\frac{1}{\tau_{0}}{R^{2}k^{2}}
\int _0^{ \pi }d\varphi \, e^{-2Rk\left|\sin\left(\varphi/2\right)\right|}\, \sin ^2\varphi .\hskip0.5cm
\end{eqnarray}
Finally, the vertex function at
$\varepsilon=k^2/2m^{*}$  has the form:
\begin{eqnarray}
\label{a12}
J_x({\bf k}; \varepsilon ,\varepsilon )=\frac{e k_x}{m^{*}}\, \frac{\tau _{\rm tr}}{\tau }.
\end{eqnarray}


\begin{thebibliography}{99}
\bibitem{scur}
Here we do not specify the mechanism of spin current generation. The existence of spin
vector potential is related to the gauge invariance of Hamiltonian $\hat{H}_0$
with respect to local unitary transformations (rotations) in the spin space.\cite{chandra90,tatara17}
\end{thebibliography}
\end{document}